\documentclass[usenatbib]{mn2e}
\usepackage{graphicx,ifthen,url}
\newcommand {\uJy}{$\mu$Jy}
\newcommand {\um}{$\mu$m}
\newcommand {\co}{\rm CO}

\newcommand {\aap}{A\&A}
\newcommand {\apj}{ApJ}
\newcommand {\apjs}{ApJS}
\newcommand {\apjl}{ApJL}
\newcommand {\mnras}{MNRAS}

\newcommand {\nat}{Nature}
\newcommand {\araa}{ARA\&A}
\newcommand {\qjras}{QJRAS}
\def\um     {$\mu$m}
\def\ts     {\thinspace}
\def\kms    {\ifmmode{{\rm \ts km\ts s}^{-1}}\else{\ts km\ts s$^{-1}$}\fi}
\def\msol   {\ifmmode{{\rm M}_{\odot}}\else{M$_{\odot}$}\fi}
\def\lsol   {\ifmmode{{\rm L}_{\odot}}\else{L$_{\odot}$}\fi}
\def\zsol   {\ifmmode{{\rm Z}_{\odot}}\else{Z$_{\odot}$}\fi}
\def\etal   {{\rm et\ts al.}}
\def\aco    {\ifmmode{^{12}{\rm CO}(J\!=\!1\! \to \!0)}\else{$^{12}${\rm CO}($J$=1$\to$0)}\fi}
\def\bco    {\ifmmode{^{12}{\rm CO}(J\!=\!2\! \to \!1)}\else{$^{12}${\rm CO}($J$=2$\to$1)}\fi}
\def\cco    {\ifmmode{^{12}{\rm CO}(J\!=\!3\! \to \!2)}\else{$^{12}${\rm CO}($J$=3$\to$2)}\fi}
\def\dco    {\ifmmode{^{12}{\rm CO}(J\!=\!4\! \to \!3)}\else{$^{12}${\rm CO}($J$=4$\to$3)}\fi}
\def\gco    {\ifmmode{^{12}{\rm CO}(J\!=\!7\! \to \!6)}\else{$^{12}${\rm CO}($J$=7$\to$6)}\fi}
\def\gn     {\rm GN20}
\def\gnb     {\rm GN20.2}
\def\gnlong {\rm SMM\,J123711.9+622212}
\def\gnblong {\rm SMM\,J123708.8+622202}
\def\ci     {\ifmmode{{\rm C}{\rm \small I}}\else{C\ts {\scriptsize I}}\fi}
\def\hi     {\ifmmode{{\rm H}{\rm \small I}}\else{H\ts {\scriptsize I}}\fi}
\def\hh     {\ifmmode{{\rm H}_2}\else{H$_2$}\fi}
\def\cone {\ifmmode{{\rm C}{\rm \small I}(^3\!P_1\!\to^3\!P_0)}
     \else{C\ts {\scriptsize I}{\small$(^3\!P_1\!\to\,^3\!P_0)$}}\fi}
\def\ctwo {\ifmmode{{\rm C}{\rm \small I}(^3\!P_2\!\to\,^3\!P_1)}
     \else{C\ts {\scriptsize I}{\small$(^3\!P_2\!\to\,^3\!P_1)$}}\fi}
\def\cij {\ifmmode{{\rm C}{\rm \small I}\,(^3P_i\to^3P_j)}\else{C\ts {\scriptsize I}\,{\small$(^3P_i\to^3P_j)$}}\fi}
\def\cii    {\ifmmode{{\rm C}{\rm \small II}}\else{C\ts {\scriptsize II}}\fi}
\def\tex {\ifmmode{{T}_{\rm ex}}\else{$T_{\rm ex}$}\fi}
\def\tmb {\ifmmode{{T}_{\rm mb}}\else{$T_{\rm mb}$}\fi}
\def\tkin {\ifmmode{{T}_{\rm kin}}\else{$T_{\rm kin}$}\fi}
\def\microns {\ifmmode{\mu{\rm m}}\else{$\mu$m}\fi}
\def\nhh   {\ifmmode{n({\rm H}_2)}\else{$n$(H$_2$)}\fi}

\newcommand{\ltaraw}{$\; \buildrel < \over \sim \;$}
\newcommand{\lta}{\lower.5ex\hbox{\ltaraw}}
\newcommand{\gtaraw}{$\; \buildrel > \over \sim \;$}
\newcommand{\gta}{\lower.5ex\hbox{\gtaraw}}

\loadboldmathitalic 
\title [Searching for Neutral Carbon in two $z\sim4$ SMGs]
{A Search for Neutral Carbon towards two z=4.05 Submillimetre Galaxies, \gn\ and \gnb}
\author[C.~M. Casey et al.]
{
C.~M. Casey$^1$\thanks{ccasey@ast.cam.ac.uk}, S.~C. Chapman,$^1$, E. Daddi$^2$, H. Dannerbauer$^3$, A. Pope$^4$, D. Scott$^5$, 
\newauthor F. Bertoldi$^6$, R.~J. Beswick$^7$, A.~W. Blain$^8$, P. Cox$^9$, R. Genzel$^{10}$, T.~R. Greve$^3$, 
\newauthor R.~J. Ivison$^{11,12}$, T.~W.~B. Muxlow$^7$, R. Neri$^9$, A. Omont$^{13}$, I. Smail$^{14}$, L.~J. Tacconi$^{10}$\\
$^1$ Institute of Astronomy, Madingley Road, Cambridge, CB3 0HA, U.K.\\
$^2$ Laboratoire AIM, CEA/DSM-CNRS-Universit\'{e} Paris Diderot, Irfu/SAp, Orme des Merisiers, F-91191 Gif-sur-Yvette, France \\
$^3$ MPIA, K\"{o}nigstuhl 17, D-69117 Heidelberg, Germany \\
$^4$ NOAO, 950 North Cherry Ave, Tucson, AZ 85719, U.S.A. \\
$^5$ Department of Physics and Astronomy, University of British Columbia, Vancouver, BC V6T 1Z1, Canada \\
$^6$ Argenlander Institute for Astronomy, University of Bonn, Auf dem H\"{u}gel 71, 53121 Bonn, Germany \\
$^7$ Jodrell Bank Centre for Astrophysics, School of Physics and Astronomy, University of Manchester, Oxford Road, Manchester, M13 9PL, U.K. \\
$^8$ Department of Astronomy, California Institute of Technology, 1200 E California Blvd, Pasadena, CA, 91125, U.S.A. \\
$^9$ Institut de Radio Astronomie Millimetrique (IRAM), St. Martin d'Heres, France \\
$^{10}$ MPE, Giessenbachstrasse 1, D-85741 Garching, Germany \\
$^{11}$ UK Astronomy Technology Centre, Royal Observatory, Blackford Hill, Edinburgh, EH9 3HJ, U.K. \\
$^{12}$ Institute for Astronomy, University of Edinburgh, Blackford Hill, Edinburgh, EH9 3HJ, U.K. \\
$^{13}$ Institut d'Astrophysique de Paris, CNRS and Universit{\'e} Pierre et Marie Curie, 
         98 Bis Boulevard Arago, 75014 Paris, France \\
$^{14}$ Institute for Computational Cosmology, Durham University, South Road, Durham DH1 3LE, U.K. \\
}

\date{Accepted 2009 August 7.  Received 2009 August 6; in original form 2009 June 30.}
\pagerange{\pageref{firstpage}--\pageref{lastpage}} \pubyear{2009}

\begin{document} 
\maketitle 
\label{firstpage}

\begin{abstract}
  Using the IRAM Plateau-de-Bure Interferometer (PdBI) we have searched for the upper fine structure line
  of neutral carbon (\ctwo, $\nu_{\rm rest} = 809$\,GHz) and \gco\ ($\nu_{\rm rest}$=806\,GHz) towards the 
  submillimetre galaxies (SMGs) \gn\ (\gnlong, $z=4.055$) and \gnb\ (\gnblong, $z=4.051$).  The far-infrared 
  (FIR) continuum is detected at 8$\sigma$ significance in \gn, with a flux density of S$_{\rm 1.8\,mm}$\,=\,1.9$\pm$0.2\,mJy, while 
  no continuum is detected in \gnb.  Both sources are statistically undetected in both \ctwo\ and \gco\ lines;  we 
  derive line luminosity limits for both \ci\ and \co\ of $L^\prime$\lta$2\times10^{10}$\,K\,km\,s$^{-1}$\,pc$^2$.  
  Assuming carbon excitation temperatures of \tex\,=\,30\,K (the galaxies' measured dust temperatures), we infer \ci\ 
  mass limits of $M_{\rm \ci}\,<\,5.4\times10^6$\,\msol (\gn) and $M_{\rm \ci}\,<\,6.8\times10^6$\,\msol (\gnb).
  The derived \ci\ abundance limits are $<\,1.8\times10^{-5}$ for \gn\ and $<\,3.8\times10^{-5}$ for \gnb\, implying 
  that the systems have Milky Way level neutral carbon enrichment (X[\ci]/X[\hh]) or lower, similar to high-redshift
  carbon-detected systems (at $5\times10^{-5}$) but about 50 times less than the neutral carbon enrichment of 
  local starburst galaxies.  Observations of \gn\ and \gnb\ in high-resolution MERLIN+VLA radio maps of GOODS-N 
  are used to further constrain the sizes and locations of active regions.  We conclude that the physical gas 
  properties of young rapidly evolving systems like \gn\ and \gnb\ are likely significantly different than 
  starburst/ULIRG environments in the local Universe yet similar to $z\sim2$ SMGs.  Unless gravitationally 
  amplified examples can be found, observations of galaxies like \gn\ will require the order of magnitude 
  increase in sensitivity of the Atacama Large Millimetre Array (ALMA) to constrain their \ci\ and high-J 
  \co\ content, despite the fact that they are the brightest systems at $z\sim$4.
\end{abstract}

\section{Introduction}\label{introduction}

Examining the molecular gas content of distant galaxies is fundamental to 
our understanding of galaxy formation and evolution theories.  Recent 
observations have shown that many distant objects contain giant molecular 
gas reservoirs \citep[$>10^{10}$\,\msol; for a review see][]{solomon05a}.  
These gas repositories are thought to provide the fuel needed to power the 
extreme starbursts observed at high redshift through their far-IR 
luminosities, $L_{\rm FIR}\sim10^{12-13}$~L$_\odot$ \citep[e.g.][]{frayer99a,neri03a,greve05a,tacconi06a,chapman08a}.
In addition, detection of AGN in molecular gas suggests a link between
the most massive starbursts, the growth of massive black holes and the onset 
of strong nuclear activity \citep{coppin08a}.  Molecular gas has already been detected in about 
60 high redshift sources, from star forming galaxies to quasars, at redshifts
$2<z<6.4$ \citep{walter03a,bertoldi03a,greve05a,tacconi06a,tacconi08a}.

Carbon monoxide has bright rotational transitions and is therefore the most
commonly observed tracer of molecular gas; however, CO lines are usually 
optically thick and difficult to model.  In contrast, observations of optically 
thin neutral carbon (\ci) can be used to derive gas physical properties 
without requiring detailed radiative transfer models.  This is because carbon 
has a 3P fine-structure forming a simple, easily-analyzed three-level system.  
The gas excitation temperature, neutral carbon column density and mass 
may be derived independent of any other information provided there are detections of
{\it both} carbon lines, \cone\ (492\,GHz) and \ctwo\ (809\,GHz) \citep[e.g.
][]{stutzki97a,weiss03a,weiss05a}.  Furthermore, observations of either \ci\ 
line could potentially be used as a probe of internal gas distribution which 
is independent from the more luminous optically thick CO observations.

While the Earth's atmosphere has very low transmission at the rest frequency of \cone\ and is
virtually opaque at \ctwo, some studies of neutral carbon have been performed in nearby
sources where the lines are bright enough to be observed.  These include the Galactic 
Centre, molecular clouds in the galactic disk, M82 and other nearby galaxies 
\citep[e.g.][]{white94a,stutzki97a,gerin00a,ojha01a,israel02a,schneider03a,kramer04a}.  
These studies show that \ci\ and CO emission trace each other well, independent
of the type of heating environment.  Since they share similar critical densities, 
$n_{\rm cr} \sim 10^3\,{\rm cm}^{-3}$, this suggests that the \cone\ and \aco\ transitions 
arise from the same gas volume and have similar excitation temperatures \citep{ikeda02a}.
In addition, several studies have found excellent agreement between \ci\ and \co\ 
derived \hh\ masses in local ULIRGs \citep[e.g.][]{gerin98a,papadopoulos04a}.

Despite improved atmospheric observing conditions at the redshifted ($z>2$)
\ci\ frequencies, observations of distant, faint sources are particularly difficult;
neutral carbon has only been confirmed previously in five other high redshift sources, 
three at $z\sim$2.5 \citep{weiss03a,weiss05a}, one $z=3.91$ QSO \citep{wagg06a} and one 
$z=4.12$ QSO \citep{pety04a}.  While it is challenging to detect, \ci\ is a cooling line 
and therefore is important in understanding the composition of a galaxy's dense interstellar 
medium (ISM).  In high redshift sources, detection of \ci\ indicates that the ISM has condensed 
and become significantly enriched while the Universe was still very young.

Here we report on the search for the upper fine structure line of neutral carbon, \ctwo, 
as well as the \gco\ line, towards two of the highest redshift and most luminous submillimetre galaxies, 
\gn\ \citep[$z=4.055$, \gnlong, identified in Pope~\etal, 2006, whose redshift was accurately 
measured in ][ hereafter referred to as D09]{daddi09a} and \gnb\ \citep[$z=4.051$, \gnblong, 
also detected in \dco\ in D09; originally from the catalogue of ][]{chapman01a}. \gn\ 
($S_{\rm 850}$=20.3\,mJy) and \gnb\ ($S_{\rm 850}$=9.9\,mJy) are two of the brightest 
submillimetre galaxies in GOODS-N \citep[see also ][]{pope07a}.  Both of their rest-frame ultraviolet spectra lack 
any emission features (D09) and are similar to the UV spectra 
of many other $z\sim2.5$ SMGs \citep{chapman05a}.  The absence of AGN characteristics in 
optical spectra suggests that the gas is mainly heated by star formation.  \citet{pope06a} 
have also shown that \gn\ has a very similar SED and mid-IR spectral properties to other 
SMGs, as are its X-ray luminosity and photon index \citep[cf.][]{alexander05a}.

In addition to our interferometric millimetre observations from the Plateau de Bure Interferometer,
we present high-resolution radio maps of both galaxies at 1.4\,GHz from MERLIN+VLA to further
characterize these bright SMGs.  We organize the paper as follows: observations are described in
Section \ref{observations}; results are discussed in Section \ref{results}; and our discussion of
the implications on high-redshift SMG enrichment are given in Section \ref{discussion}.
Throughout, we use a $\Lambda$ CDM cosmology with $H_{\rm 0} = 71$\kms~Mpc$^{-1}$ and
$\Omega_{\rm m}=0.27$ \citep{hinshaw09a}.

\section{Observations}\label{observations}

\subsection{Molecular Line Observations}

Observations were carried out with the IRAM Plateau-de-Bure interferometer through
August 2008, with a 5 dish D-configuration (i.e. compact).  We used the 2--mm receivers 
tuned to 159.873\,GHz, midway between the expected redshifted frequencies of the \ctwo\ 
($\nu_{\rm rest}=809.342$~GHz) and \gco\ ($\nu_{\rm rest}=806.651$~GHz) transitions for GN20
at $z=4.055$ and GN20.2 at $z=4.051$.  The pointing centre was closer to \gn\ than to \gnb; 
since they are both off phase center, their fluxes require primary beam attenuation
correction factors of 1.11 (\gn) and 2.62 (\gnb).  This observation was made possible because 
of the redshift measurement from the \dco\ D09 observation.  The synthesized beam size at 
160\,GHz is $\approx\,$3\arcsec.  Calibration was obtained every 12\,min using the standard 
hot/cold--load absorber measurements.  The source $3{\rm C}454.3$ was used for absolute flux 
calibration.  The antenna gain was found to be consistent with the standard value of 
29\,Jy\,K$^{-1}$ at 160\,GHz.  We estimate the flux density scale to be accurate to 
about $\pm15$\%.

Data were recorded using both polarizations overlapping, covering a 900\,MHz bandwidth.
The total on-source integration time was 8\,hrs.  The data were processed using the {\sc 
GILDAS} packages {\sc CLIC} and {\sc MAPPING} and analyzed with our own IDL-based routines.  
The extracted data cube (2 sky coordinate axes and 1 spectral axis) has an RMS noise of 
0.6\,mJy.  For clarity of presentation, we have re-gridded the data to a velocity resolution 
of $\sim$37\,\kms\ (20~MHz).  No obvious ($>$5$\sigma$) emission lines are seen in the data;
however, the next section statistically tests for the presence of FIR continuum and \co\ and 
\ci\ emission lines.

\begin{table}
\begin{center}
\caption{Observed and Derived Properties of \gn\ and \gnb}
\label{tab1}
\begin{tabular}{lrr}
\hline\hline
 & \gn\ & \gnb\ \\
\hline
RA$_{\rm MERLIN+VLA}$ (J2000)   & {\sc 12:37:11.96} & {\sc 12:37:08.80} \\
Dec$_{\rm MERLIN+VLA}$ (J2000)   & {\sc +62:22:12.4} & {\sc +62:22:01.9} \\
RA$_{\rm 160}$ (J2000)   & {\sc 12:37:11.89} & ... \\
Dec$_{\rm 160}$ (J2000)   & {\sc +62:22:12.4} & ... \\
$S_{\rm 1.4}$$^a$ (\uJy)       & 73.8$\pm$12.8     & 170.0$\pm$12.8 \\
$S_{\rm 3mm}$$^a$ (mJy)            &  0.33 & $<$0.2 \\
$S_{\rm 1.2mm}$$^a$ (mJy)            & 9.3 & $<$2.7 \\
$S_{\rm 1.1mm}$$^a$ (mJy)            & 11.5 & ... \\
$S_{\rm 850}$$^a$ (mJy)            & 20.3 & 9.9 \\
$R_{1/2}$ MERLIN+VLA (\arcsec) & 0.38$\pm$0.15 & 0.30 \\
$\beta$                       & 1.4 (derived)         & 1.5 (fixed)       \\  
$L_{FIR}$ (L$_\odot$)           & 1.0$\times$10$^{13}$ & 5.0$\times$10$^{12}$ \\
$T_{\rm dust}$  (K)         & 30$\pm$4    &     30$\pm$12 \\
$z_{\rm CO[4-3]}$$^{b}$                     & 4.055$\pm$0.001 & 4.051$\pm$0.003 \\
$I_{\rm CO[4-3]}$$^{b}$ (Jy\,km\,s$^{-1}$)     & 1.5$\pm$0.2 & 0.9$\pm$0.3 \\
$L^\prime_{\rm CO[4-3]}$$^b$ (K\,km\,s$^{-1}$\,pc$^2$) & 6.2$\times$10$^{10}$ & 3.7$\times$10$^{10}$ \\
$M_{\rm \hh}$$^{b,c}$ (M$_\odot$)             & 5.0$\times$10$^{10}$  & 3.0$\times$10$^{10}$ \\
$S_{\rm 160}$ Continuum (mJy)               & 1.9$\pm$0.2 & 0.5$\pm$0.3 \\
$I_{\rm CO[7-6]}$\,=\,$I_{\rm \ci}$ (Jy\,km\,s$^{-1}$)$^d$        & $<$\,1.2 & $<$\,1.9 \\
$L^\prime_{\rm CO[7-6]}$\,=\,$L^\prime_{\rm \ci}$ (K\,km\,s$^{-1}$\,pc$^2$)  & $<\,1.6\,\times10^{10}$ & $<\,2.5\times10^{10}$ \\
$M_{\rm \hh}$ (from CO[7-6]) (M$_\odot$)       & $<\,1.3\times10^{10}$ & $<\,2.0\times10^{11}$ \\
$M_{\rm \ci}\vert_{\rm \tex}$ (M$_\odot$)   & $<\,5.4\times10^6$ & $<\,6.8\times10^6$ \\
$X[\ci]/X[\hh]$                      & $<\,1.8\times10^{-5}$ & $<\,3.8\times10^{-5}$ \\
\hline\hline
\end{tabular}
\end{center}
{\small
{\bf Table Notes.} 
We fit modified blackbodies to the FIR continuum flux densities (see section \ref{sec:continuum}) 
to derive dust temperature and $L_{\rm FIR}$.  $\beta$ is reliably constrained to 1.4 for \gn\ and
fixed to 1.5 for \gnb.

$^{a}$ Observed continuum flux densities from the literature at 850\um\ \citep{pope06a}, 1.1mm \citep{perera08a}, 
1.2mm \citep{greve08a}, and 3mm (D09).  Upper limits for \gnb\ are 2$\sigma$.  The radio flux ($S_{\rm 1.4}$) is measured
at 1.4\,GHz from VLA.

$^{b}$ Observed properties from D09, given here for comparison.  

$^{c}$ Derived from the measured D09 CO[4-3] line strength.  D09 assumes a constant brightness temperature between \aco\ and \dco\ to derive M$_{\hh}$, 
but here we assume that $L^\prime_{\rm CO[4-3]}/L^\prime_{\rm CO[1-0]}\,=\,8/16$ \citep[cf.][]{weiss07a}, and 
$L^\prime_{\rm CO[7-6]}/L^\prime_{\rm CO[1-0]}\,=\,8/49$ \citep[cf.][]{weiss07a}. We also assume
a conversion factor $X$ = M$_{H_2}$/L$'_{CO}$\,=\,0.8~M$_\odot$~(K~km~s$^{-1}$~pc$^2$)$^{-1}$ \citep[cf.][]{downes98a}.

$^{d}$ Line intensity limits are 2$\sigma$.  In this data, both I$_{\rm \co}$ and I$_{\rm \ci}$ limits are equal since the
noise measurement in each frequency range is equal, as are the assumed line widths (700\,\kms).
}
\end{table}

\subsection{MERLIN+VLA Radio Imaging}

\begin{figure}
\centering
\includegraphics[width=0.99\columnwidth]{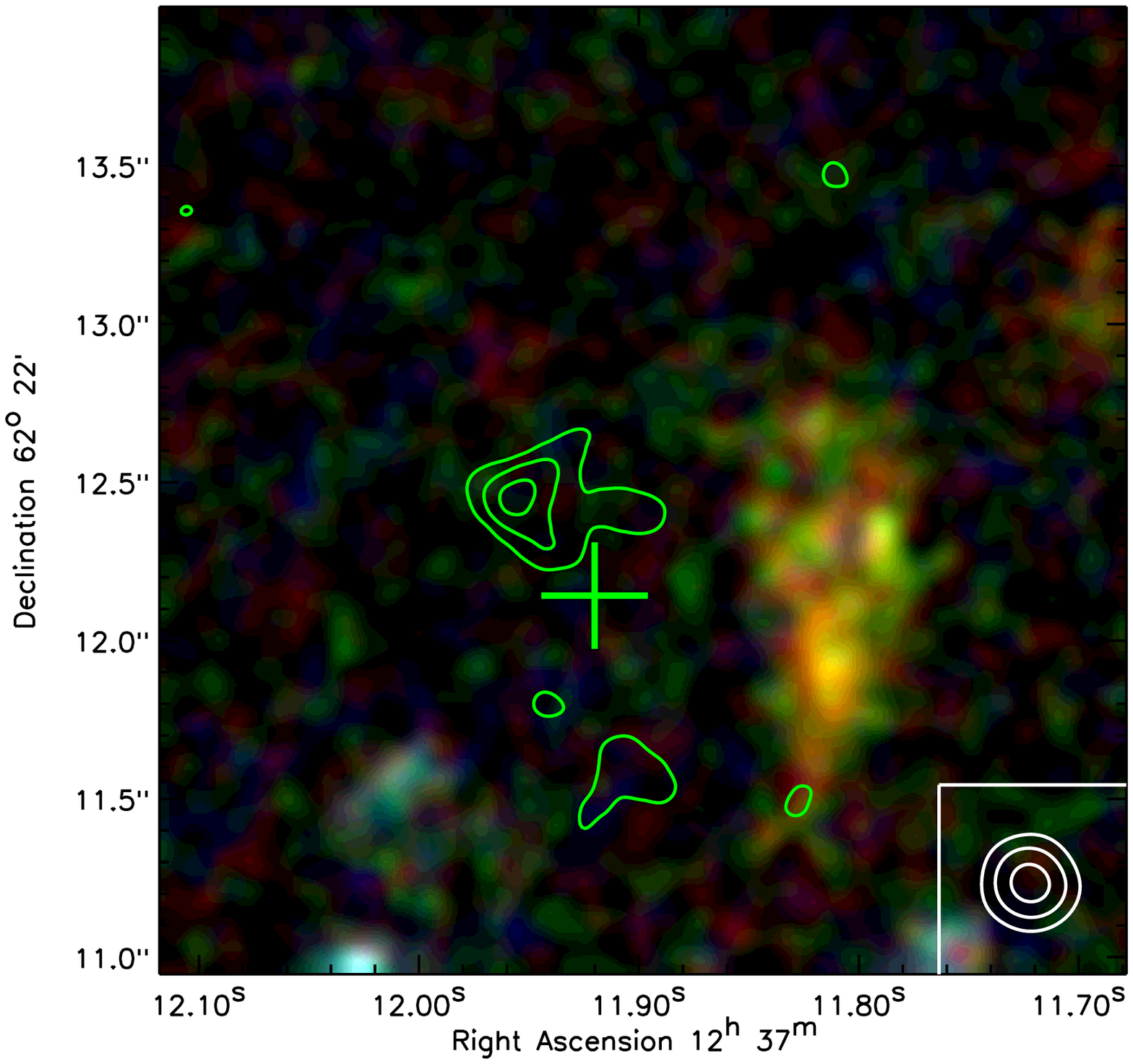}\\
\includegraphics[width=0.99\columnwidth]{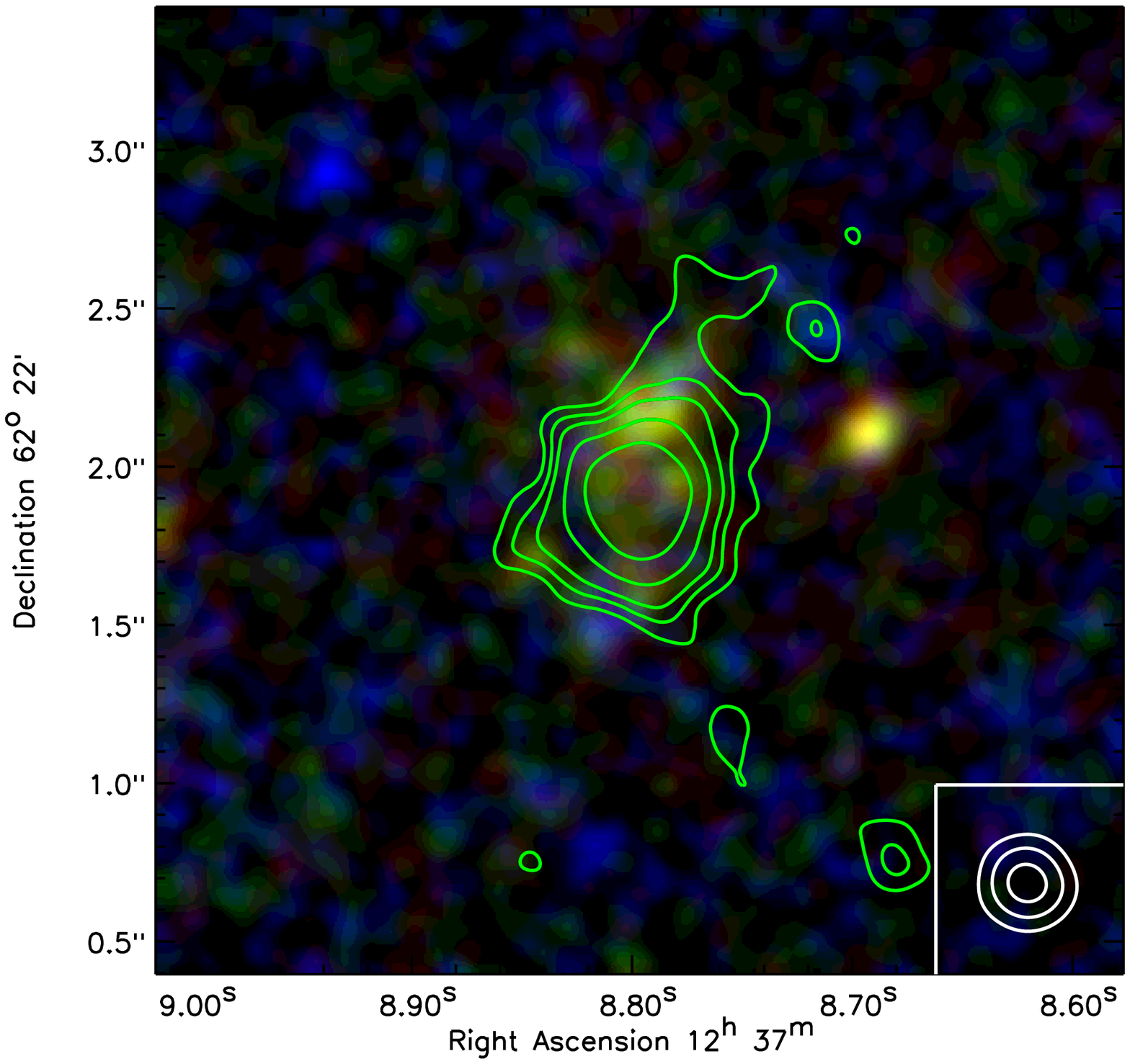}
\caption{MERLIN+VLA radio contour overlay on $HST$/ACS $BVi$ colour images of \gn\ and \gnb. The field sizes are
  3$''\times3''$ and the MERLIN beam size is 0.3$''$ (outer contour of corner inset is the FWHM of the beam).  The levels 
  of the radio contours are drawn at 3, 4, 5, 7, and 10\,$\sigma$; the RMS noise in the MERLIN+VLA images is
  2.9\,\uJy\ and 4.6\,\uJy\ for \gn\ and \gnb\ respectively.  The centre of the SMA continuum emission 
  analyzed by \citet{iono06a} and \citet{younger08a} is marked by the green cross on the \gn\ map.
  The emission centroid for \gn\ in the mid-IR ($Spitzer$ IRAC/MIPS) is consistent with the radio 
  and FIR continuum position.  The FIR-optical offset for \gn\ is significant at the 3\,$\sigma$ level.}
\label{fig:merlin}
\end{figure}

High-resolution observations from the Multi-Element Radio Linked Interferometer Network 
\citep[MERLIN;][]{thomasson86a} were obtained for these sources as described in \citet{muxlow05a},
with an RMS noise of $\sim5$\,\uJy\,beam$^{-1}$.  While \gn\  or \gnb\ were not imaged directly in 
\citet{muxlow05a}, the source is within the sensitive region of the data (Lovell Telescope 
primary beam). A combined MERLIN+VLA map was then constructed, with a sensitivity of 
3-4\,\uJy\,beam$^{-1}$ and high positional accuracy (tens of mas).  In Figure \ref{fig:merlin}, we 
show MERLIN radio contours on top of the $HST$ $BVi$ tricolor images \citep{giavalisco04a}\footnote{
Based on observations made with the NASA/ESA {\it Hubble Space Telescope}, and obtained from the Hubble 
Legacy Archive, which is a collaboration between the Space Telescope Science Institute (STScI/NASA), 
the Space Telescope European Coordinating Facility (STECF/ESA) and the Canadian Astronomy Data Centre 
(CADC/NRC/CSA).}. The field size is 3$''\times3''$ and the restoring beam size is 0.3$''$. 
We measure the center of MERLIN+VLA radio emission as {\sc 12:37:11.96, +62:22:12.4} for \gn\ and 
{\sc 12:37:08.80, +62:22:01.9} for \gnb.

\section{Results}\label{results}

\subsection{The Continuum Contribution at 1.88mm}\label{sec:continuum}

\begin{figure}
\centering
\includegraphics[width=0.99\columnwidth]{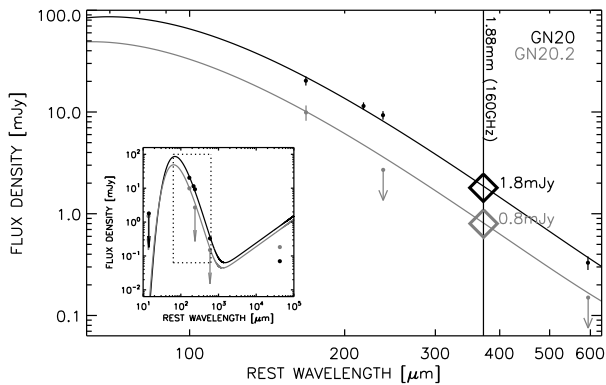}
\caption{The SED fits to the FIR measured photometry of \gn\ (black) and \gnb\ (gray).  The
inferred continuum flux densities at 1.88\,mm are shown as large diamonds: 1.8$\pm$0.4\,mJy for
\gn\ and 0.8$\pm$0.4\,mJy for \gnb.  Their full SEDs from 10\,\um\ near-IR to 10cm radio 
(rest wavelength) are shown in the inset; the area outlined in the dotted box is the zoomed-in region 
shown in the larger plot.}
\label{fig:continuum_plot}
\end{figure}

Interpreting the detectability and significance of emission lines in the millimetre requires 
an estimation of the expected continuum contribution at the wavelength of observations 
(160\,GHz, which is a wavelength of $\sim$1.88\,mm).  Independent of our PdBI observations, 
we estimate the continuum flux at 1.88\,mm by fitting modified-blackbody SEDs to the galaxies' 
FIR flux measurements.  The continuum flux densities for \gn\ and \gnb\ are given in Table~1.
We fixed the emissivities of the galaxies 
to $\beta=1.4$ for \gn\ \citep[see ][]{pope07a} and $\beta=1.5$ for \gnb\ is common practice for
galaxies in which $\beta$ is not well constrained, although we note that letting $\beta$ vary 
has a minimal effect, $<$10\%, on the continuum flux). 
With these SED constraints we measure the dust temperatures and FIR luminosities that are given
in Table~1.
The SED fits to both galaxies may be seen in Fig.~\ref{fig:continuum_plot}.  From these 
fits we infer that the 1.88\,mm flux generated by FIR continuum should be 1.8$\pm$0.4\,mJy and 
0.8$\pm$0.4\,mJy for \gn\ and \gnb, respectively.

Having inferred the expected continuum level we must now {\it measure}
the continuum from the 160\,GHz PdBI observations.  A 1-D spectrum for each object is
first extracted at the peak integrated flux position, shown in Fig.~\ref{fig:comap}.
If we ignore the possible presence of emission lines, the average flux over the whole 
bandwidth is 1.9$\pm$0.2\,mJy for \gn\ and 0.5$\pm$0.3\,mJy for \gnb.  The latter is 
consistent with the expected continuum flux, 0.8$\pm$0.4\,mJy, for \gnb, therefore we 
conclude that neither \ctwo\ nor \gco\ is detected in \gnb.  The \gn\ average flux
is consistent with the expected continuum level, 1.8\,mJy, but since there is a large 
uncertainty in the expected flux at 1.88\,mm, it needs to be further examined to see if 
the flux excess is due to the partial detection of \ci\ and \co\ lines.

\subsection{Placing limits on \ctwo\ and \gco}

\begin{figure*} 
\centering
\includegraphics[width=0.85\columnwidth]{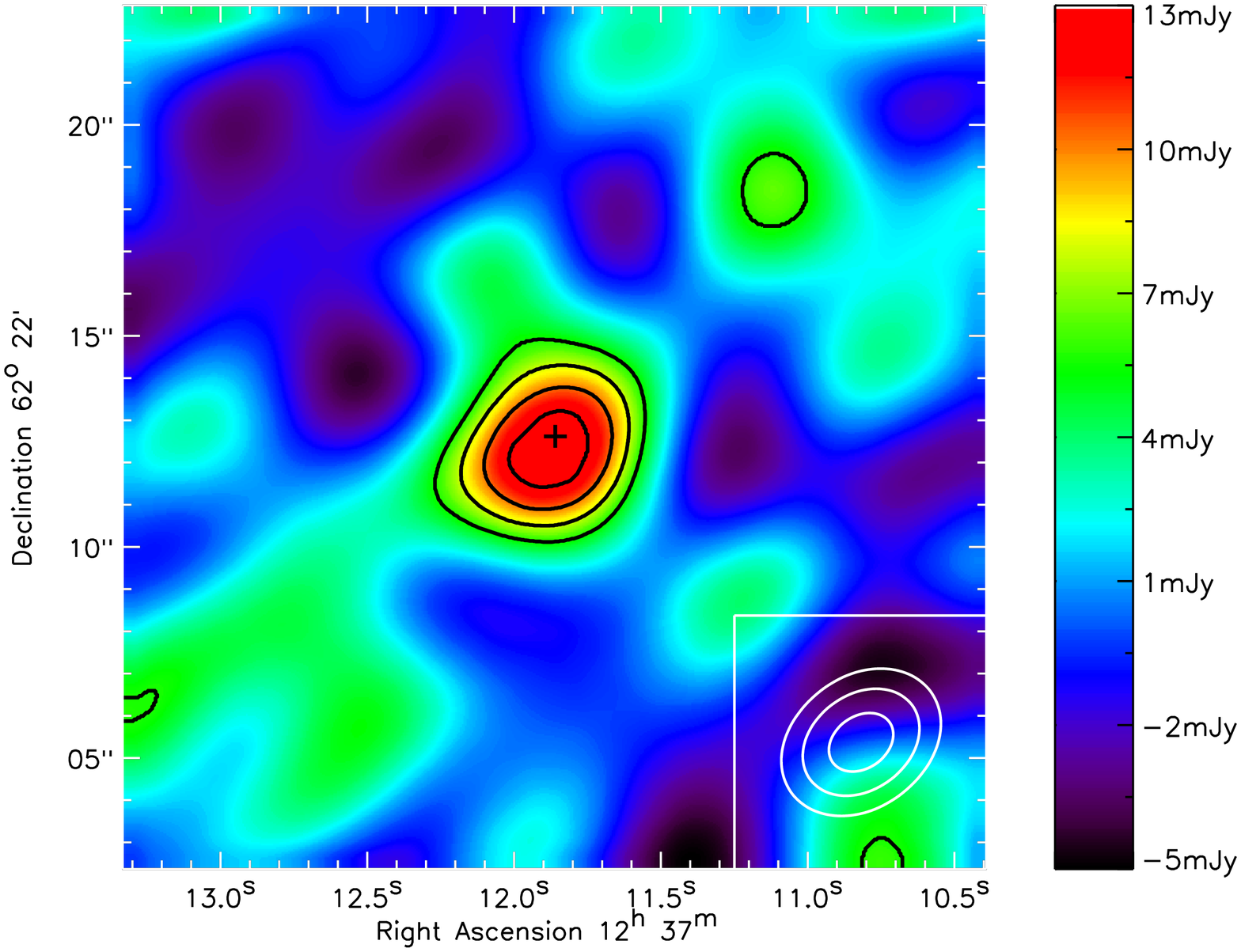}
\includegraphics[width=0.85\columnwidth]{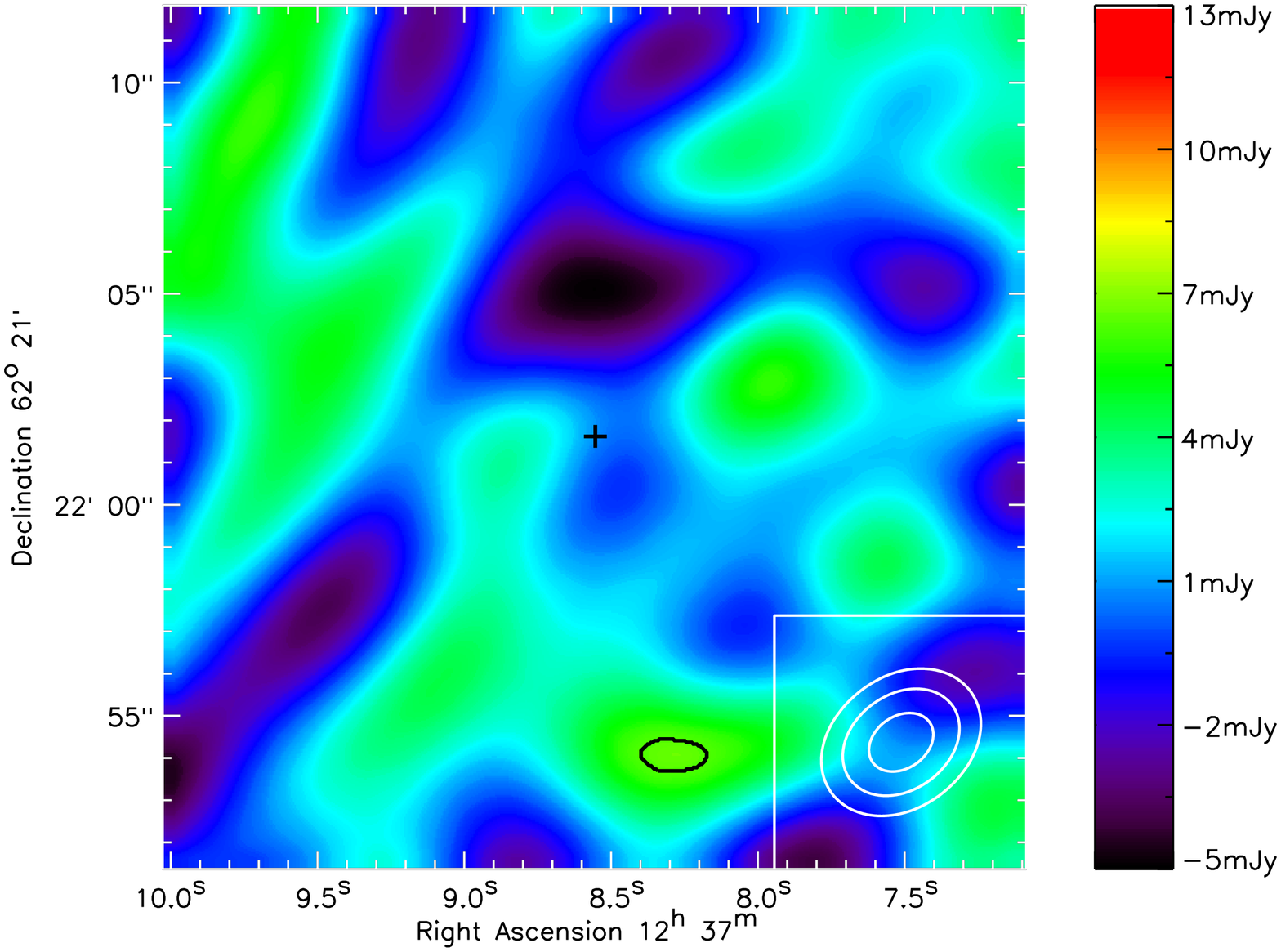}
\\
\includegraphics[width=0.99\columnwidth]{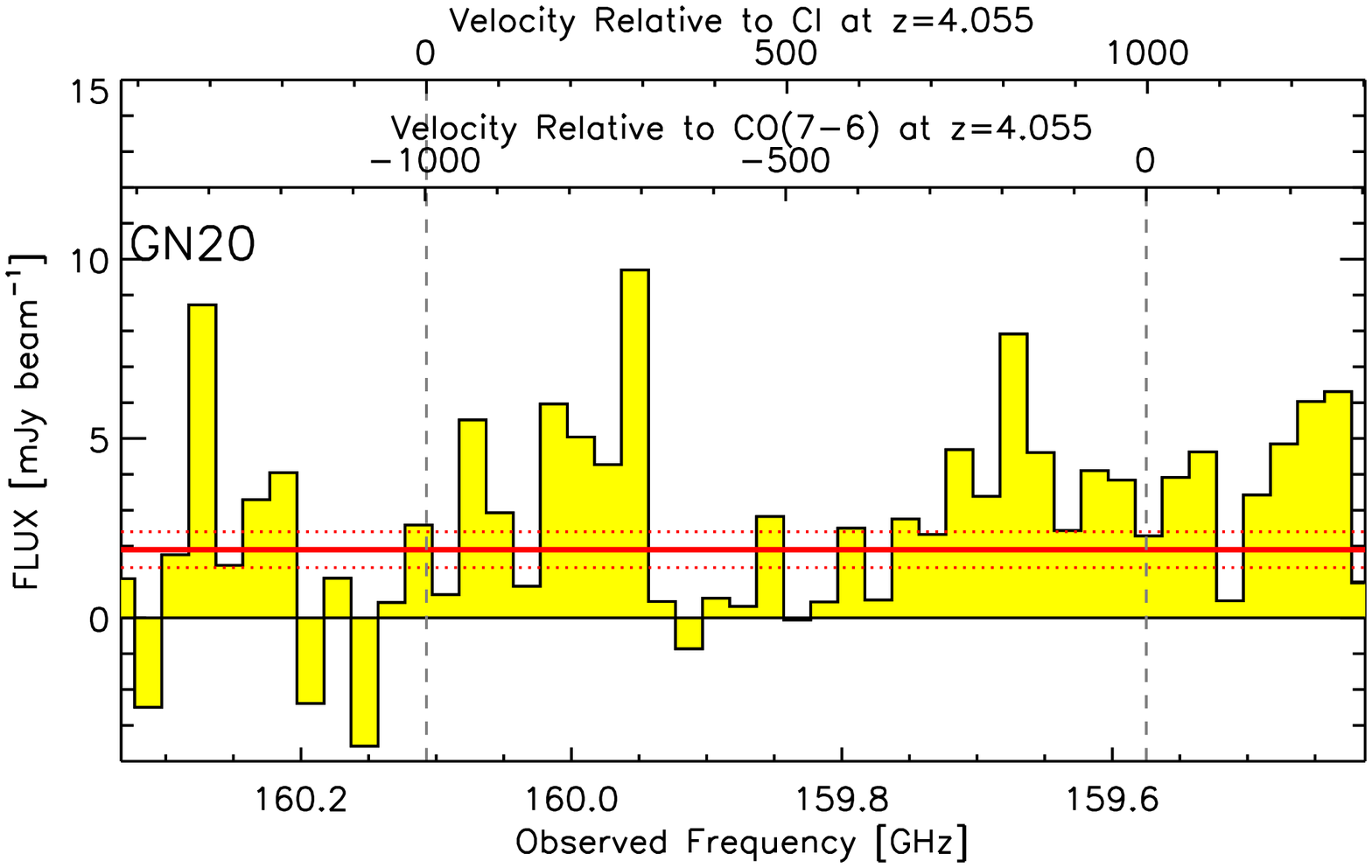}
\includegraphics[width=0.99\columnwidth]{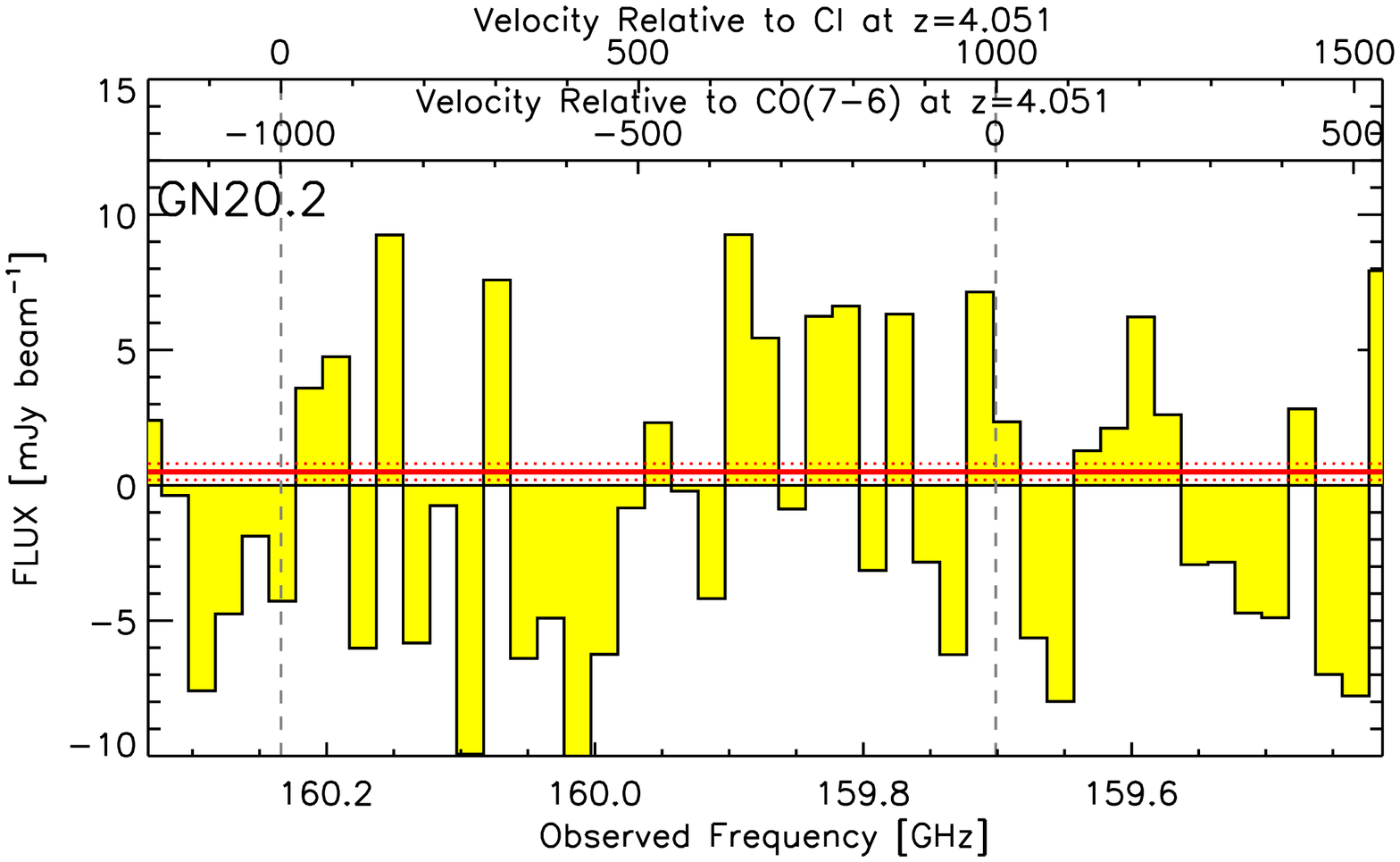}
\\
\caption{
The 160\,GHz maps and spectra for \gn\ (left) and \gnb\ (right).  The 20\arcsec\,$\times$\,20\arcsec maps 
are integrated over the whole bandwidth; 4, 5, 6, and 7\,$\sigma$ contours are over-plotted (the detection
of \gn\ is 8\,$\sigma$).  The $\sim$3\arcsec\ beam shape is shown as a lower right insert in each map.
The 20\,MHz spectra are extracted at the peak flux position of \gn\ and at the VLA centroid position for \gnb.
The average flux, or continuum estimation in the absence of line emission, is shown as a horizontal red line 
bounded by 1$\sigma$ uncertainties (red dotted lines).  The expected frequency centres of the \ci\ and \co\ 
emission lines (based on the D09 redshift) are marked by dashed gray vertical lines.  
}
\label{fig:comap} 
\end{figure*}

Figure~\ref{fig:comap} shows extracted 1-D spectra alongside integrated channel maps for \gn\ and 
\gnb. The maps are centred on each galaxy's VLA radio position, and the 1-D spectra are extracted at 
the point of peak integrated flux for \gn\ and at the VLA centroid for \gnb.  Our observations have a 
$\sim$0.6\,mJy RMS noise near \gn\ and 1.6\,mJy RMS noise near \gnb across the entire 900\,MHz bandwidth.

To measure the continuum without allowing contamination from possible line emission, we must 
extract the portion of the \gn\ spectrum where line emission is anticipated.  We use the measured D09 
redshift and line width of \dco\ ($z=4.055$ and $\Delta v\sim700\,$km\,s$^{-1}$) as an {\it a priori} 
condition and we apply them to both \gco\ and \ctwo\ lines.  Unfortunately, the narrow bandwidth 
limits the ``non-line'' spectral portion to a small number of channels, specifically those with frequencies 
159.75\,$<$\,$\nu_{\rm obs}$\,$<$\,159.95\,GHz or $\nu_{\rm obs}\,>$160.25\,GHz.  We measure the 
continuum flux density in this spectral region to be $S_{160}$(\gn)\,=\,1.4$\pm$0.3\,mJy.  This 
suggests that a flux excess {\it does} exist around the expected lines, from 159.95\,$<$\,$\nu_{\rm obs}$\,
$<$\,160.25\,GHz (the \ci\ line region) and at $\nu_{\rm obs}\,<$159.75\,GHz (the \co\ line region).  
To test the significance of these excesses, we measure the mean flux density in each region and 
compare it to the ``non-line'' flux.  The average flux density is 1.6$\pm$0.3\,mJy in the \ci\ 
line region and 2.5$\pm$0.4\,mJy in the \co\ region.  Both excesses are of low significance
($<$2$\sigma$); thus, we conclude that neither \gco\ nor \ctwo\ is detected in \gn.

With no significant detection of \co\ or \ci\ lines, we attribute all flux across the band
to the FIR continuum, therefore the measured continuum flux densities are 1.9$\pm$0.2\,mJy for \gn\ and 
0.5$\pm$0.3\,mJy for \gnb.  The latter is consistent with no detected continuum.  The former
continuum measurement for \gn\ is significant at the 8$\sigma$ level.  Although some flux from the 
emission lines might contribute to the flux density measurement of the continuum, we have not accounted
for this additional uncertainty in our measurement.

We derive 2$\sigma$ upper limits to \ci\ and \co\ line intensities assuming the redshift and line width given by
D09.  For \gn, we measure I$_{\rm \ci}\,\approx\,I_{\rm \co}<\,1.2$\,Jy\,km\,s$^{-1}$ (both limits are equal 
since the assumed line widths and the noise measurements in each frequency range are equal).  For \gnb, we measure 
I$_{\rm \ci}\,\approx\,<\,1.9$\,Jy\,km\,s$^{-1}$.  The corresponding limits to line luminosity are given in 
Table~\ref{tab1}.

\subsection{CO Excitation and Brightness Temperature}

For interpreting \co\ line intensities, the assumption of constant brightness temperature has been 
widely taken $a\ priori$ in the past due to lack of relevant data.  However, many recent observations 
of multiple \co\ transitions in high redshift sources have shown it to be an unwarranted assumption 
\citep{dannerbauer09a}.  Although detailed radiative transfer models demonstrate that the \aco\ and 
\cco\ transitions are near thermal equilibrium (i.e. $L^\prime_{{\rm CO}(1-0)} \approx 
L^\prime_{{\rm CO}(3-2)}$) for the three $z\sim2.5$ sources studied \citep{weiss05a}, there are 
several AGN dominated galaxies which do not have constant brightness temperature for high-J transitions (e.g. FSC\,10214+4724, 
APM08279+5255, and the Cloverleaf Quasar).  Here we adopt the brightness temperature conversions for SMGs 
observed by \citet{weiss07a}, see their Figure 3; explicitly, we assume flux density ratios of 
S$_{\rm CO[4-3]}$/S$_{\rm CO[1-0]}$\,=\,10.0$\pm$0.8 and S$_{\rm CO[7-6]}$/S$_{\rm CO[1-0]}$\,=\,8$\pm$1. 
This implies $L^\prime_{\rm CO[7-6]}/L^\prime_{\rm CO[1-0]}\,=\,8/49$ and
$L^\prime_{\rm CO[7-6]}/L^\prime_{\rm CO[4-3]}\,=\,64/245$.  Working backwards from the D09 \dco\ detection,
this implies that \gco\ in \gn\ would have luminosity $\sim\,$1.6$\times$10$^{10}$\,K\,km\,s$^{-1}$\,pc$^2$.  
We measure a line luminosity limit of $<\,1.6\times10^{10}$\,K\,km\,s$^{-1}$\,pc$^2$.  While there are large 
uncertainties in our brightness temperature conversion assumption, this measured result for \gn\ agrees with 
the estimated population of the higher-$J$ transitions according to other high-$z$ SMGs.  Our data for \gnb\ 
are less constraining due to the greater noise; our limit on line luminosity is consistent with 
$L^\prime_{\rm CO[7-6]}\,\approx\,1.2\times10^{10}$\,K\,km\,s$^{-1}$\,pc$^2$, which is the expected luminosity 
extrapolated from the \dco\ detection.

\subsection{Radio Source Sizes}

We use high-resolution MERLIN+VLA radio maps to constrain the size of the 1.4\,GHz emission, to better
assess the contributions from AGN and starbursts.  The measurements of angular size were made using 
{\sc AIPS} software {\sc JMFIT}, deconvolving the restoring beam before attempting to derive the 
fitted Gaussian size.  Since the signal to noise is quite low in \gn, it is impossible to detect 
component extensions smaller than the width of the restoring beam.  Figure~\ref{fig:merlin} shows 
the MERLIN+VLA 1.4GHz radio maps as contours on top of $HST$ ACS $i$-band imaging.  We measure the 
angular FWHM of the \gn\ emission region to be 0.38$\pm$0.15\arcsec, but it is only significant above 
the 0.3\arcsec\ beam size at the 2.4$\sigma$ level, while the 0.5\arcsec\ radio-optical offset is 
significant at the 3$\sigma$ level.  

\citet{iono06a} showed that the 850\,\um\ Smithsonian Millimeter Array (SMA)
position of \gn\ was significantly  offset from the optical emission (0.8\arcsec\ 
offset also at the 3$\sigma$ level).  \citet{younger08a} obtained higher 
resolution continuum imaging of \gn\ with the SMA and constrained the size 
to $0.60\pm0.13$\arcsec\ claiming it to be partially resolved.  This is consistent 
with our MERLIN+VLA-measured size of 0.38$\pm$0.15\arcsec\ due to the large uncertainties
on both measurements. Because \gn\ is not exceptionally bright in the radio (74\,\uJy), we cannot determine
if its asymmetric, extended MERLIN+VLA morphology is significant; therefore, we cannot rule out
that \gn\ might be a radio point source dominated by AGN activity.

The radio source size of \gnb\ is measured to be entirely unresolved at 
$<0.30$\arcsec.  Radio emission with $S_{1.4}\,>\,100\,$\uJy\ at $z\,=\,4.051$ 
constrained by $R_{1/2}<0.30$\arcsec\ would have to emit at a super-Eddington rate
if it were completely dominated by star formation \citep[a calculation based on the 
theoretical maximum star formation density, cf.][]{elmegreen99a,casey09a}.  For this reason, 
\gnb's radio emission is likely partially generated, if not dominated by an AGN at 
the galaxy's core.

\section{Discussion}\label{discussion}

\subsection{Neutral Carbon and \hh\ Mass}

Using the formulae given in \citet{weiss03a} for the upper fine structure 
line of neutral carbon \ctwo, we can derive the mass of neutral carbon via
\begin{equation}
\label{eqmci}
M_{\rm \ci} = C\,m_{\rm \ci}\,\frac{8 \pi k \nu_{0}^{2}}{h c^3A_{10}}\,Q(\tex)\,\frac{1}{5}\,{\rm e}^{-T_{2}/\tex}\,L^\prime_{\rm \ctwo},
\end{equation}
where $Q(\tex)=1+3{\rm e}^{-T_{1}/\tex}+5{\rm e}^{-T_{2}/\tex}$ is the
neutral carbon partition function\footnote{We note that the negative sign in the exponent
of 1/5 ${\rm e}^{-T_2/\tex}$ was accidentally omitted from the text of \citet{weiss03a} 
and \citet{weiss05a}.}.
In the equation above, $T_{1}$\,=\,23.6\,K and $T_{2}$\,=\,62.5\,K represent the energies above
the ground state, $L^\prime_\ctwo$ is given in K~\kms~pc$^2$, $m_{\rm \ci}$ is the mass of a single carbon 
atom, and $C$ is the conversion between cm$^2$ and pc$^2$, 9.5$\times$10$^{36}$~cm$^2$/pc$^2$. This
assumes both carbon lines are in local thermodynamical equilibrium (LTE) and that \ci\ emission
is optically thin.  Inserting numerical values in Equation \ref{eqmci} yields
\begin{equation}
\label{nrmci}
M_{\rm \ci} = 4.566\times10^{-4}\,Q(\tex)\,\frac{1}{5}\,e^{-62.5/ \tex}\,L^\prime_{\rm \ctwo} [\msol].
\end{equation}
Determining M$_\ci$ definitively requires measurement of the excitation temperature, which
can only be done by detecting both \cone\ and \ctwo, the two \ci\ fine structure lines.  
If $\tex$ cannot be measured directly, it is often assumed to be equal to the 
characteristic dust temperature (although in the case of the Cloverleaf Quasar, Weiss~et~al. 2003
showed that $\tex$=30\,K, significantly less than the dust temperature, 50\,K).
As the excitation temperature is unknown for both \gn\ and \gnb, we assume that $\tex\,=\,T_{\rm dust}$,
which is appropriate, since here we are only able to derive upper limits on carbon mass for both systems.  
If the excitation temperature is less than the dust temperatures (like the Cloverleaf), the upper limits
on carbon mass would decrease.  Without accounting for the possible magnification proposed by \citet{pope07a}, 
we derive upper limits on carbon mass of $5.4\times10^6\,$M$_\odot$ (\gn) and $6.8\times10^6\,$M$_\odot$ (\gnb).

Table~\ref{tab1} also lists the limits on \hh\ mass derived from the \gco\ line luminosity limits, 
assuming brightness temperature conversions from \citet{weiss07a}.  We also 
assume the standard ULIRG conversion factor $X\,=\,M_{\rm \hh}/L^\prime_{\rm \co}\,=\,$ 0.8\,M$_\odot$\,
(K\,km\,s$^{-1}$\,pc$^2$)$^{-1}$ \citep{downes98a}.  

\subsection{\ci\ Abundance in the Molecular Gas Reservoir}

Neutral carbon abundance, which is independent of magnification and adopted cosmology, is 
estimated using the ratio of masses between \ci\ and \hh\ through $X[\ci]/X[\hh]\,=\,M_\ci/(6\,M_\hh)$. 
We calculate a carbon abundance upper limit using the $M_{\rm \hh}$ value derived by D09.  We find that 
$X[\ci]/X[\hh]$\lta$\,1.8\times10^{-5}$ for \gn\ and $X[\ci]/X[\hh]$\lta$\,3.8\times10^{-5}$ for 
\gnb.  Both limits are consistent with the \ci\ abundance in the Milky Way Galaxy, 
$X[\ci]/X[\hh]=2.2\times10^{-5}$ \citep{frerking89a}, but are much lower than the measured carbon 
abundances of local starburst galaxies $\sim\,1.5\times10^{-3}$ \citep[$\sim$50 times the abundance 
of the MW, e.g.][]{schilke93a,white94a}.  \citet{israel01a} and \citet{israel03a} find similar carbon
abundances in local ULIRGs although they account for total carbon mass instead of \ci\ alone.  In comparison, 
the handful of carbon abundance measurements that have been made at high redshift average to 
$\sim\,5\pm3\times10^{-5}$ \citep[e.g.][]{weiss05a,pety04a}.  This is consistent with both the 
Milky Way's abundance and the two limits we have derived for \gn\ and \gnb.  We note that an 
abundance near (or exceeding) that of the Milky Way implies that the cold molecular gas in \gn\ 
or \gnb\ would already be significantly enriched 1.6 Gyrs after the Big Bang (a possibility that 
is not excluded by our limits).  More observations of \ci\ and \co\ in high redshift galaxies and 
QSOs are needed to investigate possible \ci\ abundance variations with redshift or variations between 
differently heated environments (star formation dominant versus AGN).

\section{Conclusions}\label{conclusions}

\gn\ has become one of the best studied high redshift starburst systems in the 
GOODS-N field, due to its high apparent infrared luminosity.  Its diverse 
multiwavelength coverage has revealed very low apparent levels of AGN emission 
relative to its high star formation.  Observations of \gn\ in molecular emission 
lines have allowed a characterization of the system's gas properties, from the strong detection 
of \dco\ of D09 to the non-detections of \ctwo\ and \gco\ in this paper. We have 
also characterized the molecular gas of \gnb, a nearby SMG at $z=4.051$.  

Millimetre and Sub-millimetre continuum observations suggest a continuum flux density at 160\,GHz 
of $S_{160\,GHz}$\,=\,1.8\,mJy for \gn\ and 0.8\,mJy for \gnb.  We measure 
1.9$\pm$0.2\,mJy continuum for \gn\ and 0.5$\pm$0.3\,mJy continuum emission from \gnb.  The
former is significant at the 8$\sigma$ level while the latter is not statistically different
from zero.

\gn\ and \gnb\ are undetected in \ctwo\ and \gco\ using observations from the IRAM-PdBI in the 
D-configuration at 160~GHz.  Line intensity upper limits are given as $I_{\rm \co[7-6]}(\gn)$=
$I_{\rm \ci}(\gn)\,<\,1.2$\,Jy\,km\,s$^{-1}$, and $I_{\rm \co[7-6]}(\gnb)$=
$I_{\rm \ci}(\gnb)\,<\,1.9$\,Jy\,km\,s$^{-1}$.

High-resolution radio imaging from MERLIN+VLA show that the radio emission of \gn\ is extended
over 0.38$\pm$0.15\arcsec\ (2.7\,kpc).  This size is consistent with the FIR continuum size 
measured by \citet{iono06a} and \citet{younger08a}.  In contrast, \gnb\ is unresolved at 
$<$0.30\arcsec\ (2.0\,kpc), suggesting that AGN emission may dominate its radio emission.

The conditions for metal enrichment of the ISM of \gn\ and \gnb\ do not appear to 
differ greatly with other high redshift sources or the Milky Way.  However, local
starburst galaxies have $\sim$50 times the carbon abundance of the high-$z$ systems
and the upper limits for \gn\ and \gnb.

Future observations from the Atacama Large Millimeter Array (ALMA) of cooling gas emission 
lines like \cone\ and \ctwo\ in high redshift sources will significantly advance the understanding 
of the relationship between early star formation and metal enrichment. Although difficult (in the 
absence of a boost from gravitational lensing) to observe 
with current instruments as evidenced by the lack of detection in \gn, one of the brightest known
high redshift SMGs, \ci\ is an excellent tracer of cold molecular gas in galaxies. It may even 
provide useful constraints on gas content when undetected relative to \co, and may be used as a 
powerful diagnostic of galaxy evolution out to the highest observable redshifts, especially in 
the light of the next generation of mm and submm telescopes.

\section*{Acknowledgments}
Based on observations carried out with the IRAM Plateau de Bure Interferometer.
IRAM is supported by INSU/CNRS (France), MPG (Germany) and IGN (Spain).
We acknowledge the use of GILDAS software ({\tt http://www.iram.fr/IRAMFR/GILDAS}).
CMC thanks the Gates-Cambridge Trust and IRS thanks STFC for support. ED gratefully
acknowledges funding support from ANR-08-JCJC-0008.

\label{lastpage}

\begin{thebibliography}{}

\bibitem[\protect\citeauthoryear{{Alexander}, {Bauer}, {Chapman}, {Smail},
  {Blain}, {Brandt} \& {Ivison}}{{Alexander} et~al.}{2005}]{alexander05a}
{Alexander} D.~M.,  {Bauer} F.~E.,  {Chapman} S.~C.,  {Smail} I.,  {Blain}
  A.~W.,  {Brandt} W.~N.,    {Ivison} R.~J.,  2005, \apj, 632, 736

\bibitem[\protect\citeauthoryear{{Bertoldi} et~al.,}{{Bertoldi}
  et~al.}{2003}]{bertoldi03a}
{Bertoldi} F.,  et~al., 2003, \aap, 409, L47

\bibitem[\protect\citeauthoryear{{Casey}, {Chapman}, {Muxlow}, {Beswick},
  {Alexander} \& {Conselice}}{{Casey} et~al.}{2009}]{casey09a}
{Casey} C.~M.,  {Chapman} S.~C.,  {Muxlow} T.~W.~B.,  {Beswick} R.~J.,
  {Alexander} D.~M.,    {Conselice} C.~J.,  2009, \mnras, 395, 1249

\bibitem[\protect\citeauthoryear{{Chapman}, {Blain}, {Smail} \&
  {Ivison}}{{Chapman} et~al.}{2005}]{chapman05a}
{Chapman} S.~C.,  {Blain} A.~W.,  {Smail} I.,    {Ivison} R.~J.,  2005, \apj,
  622, 772

\bibitem[\protect\citeauthoryear{{Chapman} et~al.,}{{Chapman}
  et~al.}{2008}]{chapman08a}
{Chapman} S.~C.,  et~al., 2008, \apj, 689, 889

\bibitem[\protect\citeauthoryear{{Chapman}, {Richards}, {Lewis}, {Wilson} \&
  {Barger}}{{Chapman} et~al.}{2001}]{chapman01a}
{Chapman} S.~C.,  {Richards} E.~A.,  {Lewis} G.~F.,  {Wilson} G.,    {Barger}
  A.~J.,  2001, \apjl, 548, L147

\bibitem[\protect\citeauthoryear{{Coppin} et~al.,}{{Coppin}
  et~al.}{2008}]{coppin08a}
{Coppin} K.~E.~K.,  et~al., 2008, \mnras, 389, 45

\bibitem[\protect\citeauthoryear{{Daddi} et~al.,}{{Daddi}
  et~al.}{2009}]{daddi09a}
{Daddi} E.,  et~al., 2009, \apj, 694, 1517

\bibitem[\protect\citeauthoryear{{Dannerbauer}, {Daddi}, {Riechers}, {Walter},
  {Carilli}, {Dickinson}, {Elbaz} \& {Morrison}}{{Dannerbauer}
  et~al.}{2009}]{dannerbauer09a}
{Dannerbauer} H.,  {Daddi} E.,  {Riechers} D.~A.,  {Walter} F.,  {Carilli}
  C.~L.,  {Dickinson} M.,  {Elbaz} D.,    {Morrison} G.~E.,  2009, \apjl, 698,
  L178

\bibitem[\protect\citeauthoryear{{Downes} \& {Solomon}}{{Downes} \&
  {Solomon}}{1998}]{downes98a}
{Downes} D.,  {Solomon} P.~M.,  1998, \apj, 507, 615

\bibitem[\protect\citeauthoryear{{Elmegreen}}{{Elmegreen}}{1999}]{elmegreen99a}
{Elmegreen} B.~G.,  1999, \apj, 517, 103

\bibitem[\protect\citeauthoryear{{Frayer} et~al.,}{{Frayer}
  et~al.}{1999}]{frayer99a}
{Frayer} D.~T.,  et~al., 1999, \apjl, 514, L13

\bibitem[\protect\citeauthoryear{{Frerking}, {Keene}, {Blake} \&
  {Phillips}}{{Frerking} et~al.}{1989}]{frerking89a}
{Frerking} M.~A.,  {Keene} J.,  {Blake} G.~A.,    {Phillips} T.~G.,  1989,
  \apj, 344, 311

\bibitem[\protect\citeauthoryear{{Gerin} \& {Phillips}}{{Gerin} \&
  {Phillips}}{1998}]{gerin98a}
{Gerin} M.,  {Phillips} T.~G.,  1998, \apjl, 509, L17

\bibitem[\protect\citeauthoryear{{Gerin} \& {Phillips}}{{Gerin} \&
  {Phillips}}{2000}]{gerin00a}
{Gerin} M.,  {Phillips} T.~G.,  2000, \apj, 537, 644

\bibitem[\protect\citeauthoryear{{Giavalisco} et~al.,}{{Giavalisco}
  et~al.}{2004}]{giavalisco04a}
{Giavalisco} M.,  et~al., 2004, \apjl, 600, L93

\bibitem[\protect\citeauthoryear{{Greve} et~al.,}{{Greve}
  et~al.}{2005}]{greve05a}
{Greve} T.~R.,  et~al., 2005, \mnras, 359, 1165

\bibitem[\protect\citeauthoryear{{Greve}, {Pope}, {Scott}, {Ivison}, {Borys},
  {Conselice} \& {Bertoldi}}{{Greve} et~al.}{2008}]{greve08a}
{Greve} T.~R.,  {Pope} A.,  {Scott} D.,  {Ivison} R.~J.,  {Borys} C.,
  {Conselice} C.~J.,    {Bertoldi} F.,  2008, \mnras, 389, 1489

\bibitem[\protect\citeauthoryear{{Hinshaw} et~al.,}{{Hinshaw}
  et~al.}{2009}]{hinshaw09a}
{Hinshaw} G.,  et~al., 2009, \apjs, 180, 225

\bibitem[\protect\citeauthoryear{{Ikeda}, {Oka}, {Tatematsu}, {Sekimoto} \&
  {Yamamoto}}{{Ikeda} et~al.}{2002}]{ikeda02a}
{Ikeda} M.,  {Oka} T.,  {Tatematsu} K.,  {Sekimoto} Y.,    {Yamamoto} S.,
  2002, \apjs, 139, 467

\bibitem[\protect\citeauthoryear{{Iono} et~al.,}{{Iono}
  et~al.}{2006}]{iono06a}
{Iono} D.,  et~al., 2006, \apjl, 640, L1

\bibitem[\protect\citeauthoryear{{Israel} \& {Baas}}{{Israel} \&
  {Baas}}{2001}]{israel01a}
{Israel} F.~P.,  {Baas} F.,  2001, \aap, 371, 433

\bibitem[\protect\citeauthoryear{{Israel} \& {Baas}}{{Israel} \&
  {Baas}}{2002}]{israel02a}
{Israel} F.~P.,  {Baas} F.,  2002, \aap, 383, 82

\bibitem[\protect\citeauthoryear{{Israel} \& {Baas}}{{Israel} \&
  {Baas}}{2003}]{israel03a}
{Israel} F.~P.,  {Baas} F.,  2003, \aap, 404, 495

\bibitem[\protect\citeauthoryear{{Kramer}, {Jakob}, {Mookerjea}, {Schneider},
  {Br{\"u}ll} \& {Stutzki}}{{Kramer} et~al.}{2004}]{kramer04a}
{Kramer} C.,  {Jakob} H.,  {Mookerjea} B.,  {Schneider} N.,  {Br{\"u}ll} M.,
  {Stutzki} J.,  2004, \aap, 424, 887

\bibitem[\protect\citeauthoryear{{Muxlow} et~al.,}{{Muxlow}
  et~al.}{2005}]{muxlow05a}
{Muxlow} T.~W.~B.,  et~al., 2005, \mnras, 358, 1159

\bibitem[\protect\citeauthoryear{{Neri} et~al.,}{{Neri}
  et~al.}{2003}]{neri03a}
{Neri} R.,  et~al., 2003, \apjl, 597, L113

\bibitem[\protect\citeauthoryear{{Ojha} et~al.,}{{Ojha}
  et~al.}{2001}]{ojha01a}
{Ojha} R.,  et~al., 2001, \apj, 548, 253

\bibitem[\protect\citeauthoryear{{Papadopoulos} \& {Greve}}{{Papadopoulos} \&
  {Greve}}{2004}]{papadopoulos04a}
{Papadopoulos} P.~P.,  {Greve} T.~R.,  2004, \apjl, 615, L29

\bibitem[\protect\citeauthoryear{{Perera} et~al.,}{{Perera}
  et~al.}{2008}]{perera08a}
{Perera} T.~A.,  et~al., 2008, \mnras, 391, 1227

\bibitem[\protect\citeauthoryear{{Pety}, {Beelen}, {Cox}, {Downes}, {Omont},
  {Bertoldi} \& {Carilli}}{{Pety} et~al.}{2004}]{pety04a}
{Pety} J.,  {Beelen} A.,  {Cox} P.,  {Downes} D.,  {Omont} A.,  {Bertoldi} F.,
    {Carilli} C.~L.,  2004, \aap, 428, L21

\bibitem[\protect\citeauthoryear{{Pope} et~al.,}{{Pope}
  et~al.}{2006}]{pope06a}
{Pope} A.,  et~al., 2006, \mnras, 370, 1185

\bibitem[\protect\citeauthoryear{{Pope}}{{Pope}}{2007}]{pope07a}
{Pope} E.~A.,  2007, PhD Thesis, University of British Columbia, p.~~

\bibitem[\protect\citeauthoryear{{Schilke}, {Carlstrom}, {Keene} \&
  {Phillips}}{{Schilke} et~al.}{1993}]{schilke93a}
{Schilke} P.,  {Carlstrom} J.~E.,  {Keene} J.,    {Phillips} T.~G.,  1993,
  \apjl, 417, L67

\bibitem[\protect\citeauthoryear{{Schneider}, {Simon}, {Kramer}, {Kraemer},
  {Stutzki} \& {Mookerjea}}{{Schneider} et~al.}{2003}]{schneider03a}
{Schneider} N.,  {Simon} R.,  {Kramer} C.,  {Kraemer} K.,  {Stutzki} J.,
  {Mookerjea} B.,  2003, \aap, 406, 915

\bibitem[\protect\citeauthoryear{{Solomon} \& {Vanden Bout}}{{Solomon} \&
  {Vanden Bout}}{2005}]{solomon05a}
{Solomon} P.~M.,  {Vanden Bout} P.~A.,  2005, \araa, 43, 677

\bibitem[\protect\citeauthoryear{{Stutzki} et~al.,}{{Stutzki}
  et~al.}{1997}]{stutzki97a}
{Stutzki} J.,  et~al., 1997, \apjl, 477, L33

\bibitem[\protect\citeauthoryear{{Tacconi} et~al.,}{{Tacconi}
  et~al.}{2006}]{tacconi06a}
{Tacconi} L.~J.,  et~al., 2006, \apj, 640, 228

\bibitem[\protect\citeauthoryear{{Tacconi} et~al.,}{{Tacconi}
  et~al.}{2008}]{tacconi08a}
{Tacconi} L.~J.,  et~al., 2008, \apj, 680, 246

\bibitem[\protect\citeauthoryear{{Thomasson}}{{Thomasson}}{1986}]{thomasson86a}
{Thomasson} P.,  1986, \qjras, 27, 413

\bibitem[\protect\citeauthoryear{{Wagg}, {Wilner}, {Neri}, {Downes} \&
  {Wiklind}}{{Wagg} et~al.}{2006}]{wagg06a}
{Wagg} J.,  {Wilner} D.~J.,  {Neri} R.,  {Downes} D.,    {Wiklind} T.,  2006,
  \apj, 651, 46

\bibitem[\protect\citeauthoryear{{Walter} et~al.,}{{Walter}
  et~al.}{2003}]{walter03a}
{Walter} F.,  et~al., 2003, \nat, 424, 406

\bibitem[\protect\citeauthoryear{{Weiss}, {Downes}, {Henkel} \&
  {Walter}}{{Weiss} et~al.}{2005}]{weiss05a}
{Weiss} A.,  {Downes} D.,  {Henkel} C.,    {Walter} F.,  2005, \aap, 429, L25

\bibitem[\protect\citeauthoryear{{Weiss}, {Downes}, {Walter} \&
  {Henkel}}{{Weiss} et~al.}{2007}]{weiss07a}
{Weiss} A.,  {Downes} D.,  {Walter} F.,    {Henkel} C.,  2007, in {Baker}
  A.~J.,  {Glenn} J.,  {Harris} A.~I.,  {Mangum} J.~G.,   {Yun} M.~S.,  eds,
  From Z-Machines to ALMA: (Sub)Millimeter Spectroscopy of Galaxies Vol.~375 of
  Astronomical Society of the Pacific Conference Series, {CO Line SEDs of
  High-Redshift QSOs and Submm Galaxies}.
p.~25

\bibitem[\protect\citeauthoryear{{Weiss}, {Henkel}, {Downes} \&
  {Walter}}{{Weiss} et~al.}{2003}]{weiss03a}
{Weiss} A.,  {Henkel} C.,  {Downes} D.,    {Walter} F.,  2003, \aap, 409, L41

\bibitem[\protect\citeauthoryear{{White}, {Ellison}, {Claude}, {Dent} \&
  {Matheson}}{{White} et~al.}{1994}]{white94a}
{White} G.~J.,  {Ellison} B.,  {Claude} S.,  {Dent} W.~R.~F.,    {Matheson}
  D.~N.,  1994, \aap, 284, L23

\bibitem[\protect\citeauthoryear{{Younger} et~al.,}{{Younger}
  et~al.}{2008}]{younger08a}
{Younger} J.~D.,  et~al., 2008, \apj, 688, 59

\end{thebibliography}
\end{document}